\documentclass[preprint,pre,nobibnotes,twocolumn,10pt]{revtex4}%
\usepackage{amsfonts}
\usepackage{amsmath}
\usepackage{amssymb}
\usepackage{graphicx}%
\providecommand{\U}[1]{\protect\rule{.1in}{.1in}}

\begin{document}
\preprint{ }
\preprint{UATP/1905}
\title{Comment on "The generalized Boltzmann distribution is the only distribution in
which the Gibbs-Shannon entropy equals the thermodynamic entropy" by X. Gao,
E. Gallicchio and A.E. Roitberg [J. Chem. Phys. \textbf{151}, 034113 (2019)]}
\author{P.D. Gujrati,$^{1,2}$ }
\affiliation{$^{1}$Department of Physics, $^{2}$Department of Polymer Science, The
University of Akron, Akron, OH 44325}
\email{pdg@uakron.edu}

\begin{abstract}
The title of the paper leads to an incorrect conclusion as we show that the
equilibrium result of the paper is a special limit of a general result for
nonequilibrium systems in internal equilibrium already available in the
literature. We also point out some of the limitations of the approach taken by
the authors.

\end{abstract}
\date{\today}
\maketitle

Gao, Gallicchio and Roitberg (GGR) have suggested in their work
\cite{Roitberg} that the generalized Boltzmann distribution is the \emph{only}
distribution for which the Gibbs-Shannon entropy $\mathcal{S}$ equals the
equilibrium (EQ) thermodynamic entropy $S_{\text{eq}}$. In this form, the
result is not new as acknowledged by them for common EQ ensembles
($NVT_{0},V\mu_{0}T_{0}$ and $NP_{0}T_{0}$) \cite{Landau} that require $n=3$
independent variables;\ the suffix $0$ has been added to the fields as a
reminder that they refer to the medium $\widetilde{\Sigma}$, which is always
in EQ. (This choice of notation will become useful below when we discuss
nonequilibrium (NEQ) systems.) The generalization to arbitrary EQ ensembles
($n>3$) is trivially done; see Guggenheim \cite{Guggenheim}. Therefore, the
main contribution of GGR is their claim that the generalized Boltzmann
distribution is the \emph{only} distribution for which $\mathcal{S}$ equals
$S_{\text{eq}}$; they donot remark that $S_{\text{eq}}$ is defined up to a
constant but not $\mathcal{S}$.

The Gibbs-Shannon entropy $\mathcal{S}=-%
{\textstyle\sum\nolimits_{k}}
p_{k}\ln p_{k}$ \cite{Note}, where $k$ indexes the microstates of the system,
is commonly applied to NEQ states. Their claim, therefore, will most certainly
force the reader to incorrectly conclude that $\mathcal{S}$ is not equal to
the thermodynamic entropy $S$ in a NEQ process where $\mathcal{S}$\ is well
defined as is easily verified for a NEQ ideal gas \cite{Note} discussed by
Landau and Lifshitz \cite{Landau}. They equate $\mathcal{S}$ with $S$
\cite[see Eq. (40.7)]{Landau} as GGR do in their Postulate 2. As $S$ satisfies
the second law, they use the \emph{entropy maximization} (akin to GGR\ using
their Eq. (9) for EQ as Postulate 1; more on this later) to derive the
equilibrium distribution (Conclusion in the GGR approach). Indeed, we have
also used $\mathcal{S}=S$ to identify $S$ in our work \cite{Guj-I,Guj-entropy}%
. By using entropy maximization, we then obtain the probability distribution
(Conclusion) for a special class of NEQ macrostates said to be in
\emph{internal equilibrium} (IEQ); see below. Thus the result by GGR is a
special limit of our more general result: the Gibbs-Shannon entropy
$\mathcal{S}$ also equals the thermodynamic entropy $S$ of NEQ systems that
are in IEQ having a generalized Boltzmann distribution.

Let $\mathbf{X}=(E,V,\cdots)$ denote the set of $n$\ (extensive) observables
of the system. In EQ, $S_{\text{eq}}(\boldsymbol{X})$ is a state function of
$\mathbf{X}$ in a state space $\mathfrak{S}_{\mathbf{X}}$. Away from EQ,
$S(\mathbf{X},t)<S_{\text{eq}}(\mathbf{X})$ has an explicit time dependence
and approaches $S_{\text{eq}}(\mathbf{X})$ from below as the system approaches
EQ \cite{Guj-I,Guj-II,Guj-entropy}. The existence of $S(\mathbf{X},t)$ is
justified by the law of increase of entropy as discussed elsewhere
\cite{Guj-entropy}. It is common to use \emph{internal variables}
\cite{deGroot,Prigogine,Maugin} to justify this extra time dependence. Let
$\boldsymbol{\xi}$ denote the set of internal variables needed to account for
this $t$-dependence so that $S(\mathbf{X},t)$ can be written as a unique state
function $S(\mathbf{Z})$ in an \emph{enlarged }state space $\mathfrak{S}%
_{\mathbf{Z}}$, $\mathbf{Z}\doteq\mathbf{X}\cup\boldsymbol{\xi}$. Such a state
in $\mathfrak{S}_{\boldsymbol{Z}}$ is identified as an \emph{internal
equilibrium state} (IEQS) for which $p_{k}$ has a special form; see Eq.
(\ref{MicroProb}). States that are not in IEQ will have their entropy given by
$S(\mathbf{Z},t)<S(\mathbf{Z})$, and in time approaches $S(\mathbf{Z})$ from
below. For them, $p_{k}$ and $\mathcal{S}(\mathbf{Z},t)$ have explicit time
dependence. Evidently, $S(\mathbf{Z},t)$ must be maximized to yield
$S(\mathbf{Z})$, which then leads to Eq. (\ref{Gibbs-S}). Thus, \emph{entropy
maximizing is equivalent to Postulate 1} as asserted above.

The Gibbs fundamental relation follows from $S(\mathbf{Z})$%
\begin{equation}
dS=\boldsymbol{\lambda}\cdot d\mathbf{Z}, \label{Gibbs-S}%
\end{equation}
with $\boldsymbol{\lambda}\doteq\partial S/\partial\mathbf{Z}$; in particular,
$\partial S/\partial E\boldsymbol{=}1/T,\partial S/\partial V\boldsymbol{=}%
P/T,\cdots$, and $\partial S/\partial\boldsymbol{\xi=}\mathbf{A}/T$; here
$\mathbf{y}=(T,P,\cdots,\mathbf{A)}$\ denotes the set of \emph{fields}
(temperature, pressure, $\cdots$, and affinity) and differs from
$\mathbf{y}_{0}=(T_{0},P_{0},\cdots,\mathbf{A}_{0}=0)$ of the medium
$\widetilde{\Sigma}$ unless $\Sigma$ is in EQ. We rewrite Eq. (\ref{Gibbs-S})
as
\begin{equation}
dE\mathbf{=}TdS-T\boldsymbol{\lambda}^{E}\cdot d\mathbf{Z}^{E},
\label{Gibbs-E}%
\end{equation}
where $\mathbf{Z}^{E}\doteq\mathbf{Z}\backslash E$, and $\boldsymbol{\lambda
}^{E}=\partial S/\partial\mathbf{Z}^{E}$. For $S_{\text{eq}}(\boldsymbol{X})$,
Eq. (\ref{Gibbs-E}) reduces to Eq. (9) of GGR \cite{Roitberg}, which is simply
a consequence of $\mathcal{S}=S_{\text{eq}}$ being a state function. Thus,
\emph{Postulate 1 is a direct consequence of Postulate 2} and is not required.

A particular NEQ ensemble in $\mathfrak{S}_{\boldsymbol{Z}}$ is identified as
follows. We pick a set of fields $\mathbf{y}_{\text{f}}\subset\mathbf{y}=$
$(T,P,\cdots,\mathbf{A)}$ and a set $\mathbf{Z}_{\text{f}}^{E}\subseteq
\mathbf{Z}^{E}=(V,\cdots,\boldsymbol{\xi})$ to be held fixed (f for fixed)
such that the number of elements in $\mathbf{y}_{\text{f}}$ and $\mathbf{Z}%
_{\text{f}}^{E}$ total $n$. We select $\mathbf{Z}_{\text{f}}^{E}$ such that
$\mathbf{Y}_{\text{f}}\cap\mathbf{Z}_{\text{f}}^{E}=\emptyset$, where
$\mathbf{Y}_{\text{f}}$ denotes the set conjugate to $\mathbf{y}_{\text{f}}$.
We must set $d\mathbf{Z}_{\text{f}}^{E}=0$ in Eqs. (\ref{Gibbs-S}) and
(\ref{Gibbs-E}). We will always require that $T\in\mathbf{y}_{\text{f}}$ so
that $TdS$ is always present in Eq. (\ref{Gibbs-E}) for these ensembles.

The Hamiltonian of $\Sigma$ contains $\mathbf{Z}^{E}$ as a parameter.
Therefore, its microstate $\mathsf{m}_{k}(\mathbf{Z}^{E})$ and its energy
$E_{k}(\mathbf{Z}^{E})$ also depend on it. Let $p_{k}$ be the probability of
$\mathsf{m}_{k}(\mathbf{Z}^{E})$; however, neither $\mathsf{m}_{k}%
(\mathbf{Z}^{E})$ nor $E_{k}(\mathbf{Z}^{E})$ depend on $p_{k}$. It follows
from $E=%
{\textstyle\sum\nolimits_{k}}
p_{k}E_{k}$ that $\boldsymbol{\lambda}^{E}=$ $%
{\textstyle\sum\nolimits_{k}}
p_{k}\boldsymbol{\lambda}_{k}^{E}$, where $\boldsymbol{\lambda}_{k}^{E}%
\doteq(-1/T)\partial E_{k}(\mathbf{Z}^{E})/\partial\mathbf{Z}^{E}$. For
example, the average pressure is $P=%
{\textstyle\sum\nolimits_{k}}
p_{k}P_{k},$ where $P_{k}=-\partial E_{k}/\partial V$ is the pressure for
$\mathsf{m}_{k}(\mathbf{Z}^{E})$. However, we wish to consider the case when
each microstate's pressure is identically equal to $P\in\mathbf{y}_{\text{f}}%
$. We accomplish this by choosing $V_{k}$ for $\mathsf{m}_{k}$ such that
$\left.  \partial E_{k}/\partial V\right\vert _{V_{k}}=-P$. However, this
results in fluctuating $V_{k}$ over $\mathsf{m}_{k}$ to describe a fixed
$P$-ensemble having the average $V=%
{\textstyle\sum\nolimits_{k}}
p_{k}V_{k}$. This comment applies to any of the fields in $\mathbf{y}%
_{\text{f}}$ so that fixed fields require corresponding conjugate variables
$\mathbf{Y}_{k\text{f}}$ such as $E_{k},V_{k},\cdots,\boldsymbol{\xi}_{k}$ to
fluctuate over the microstates. The average $\mathbf{Y}_{\text{f}}$ is
determined by $\mathbf{Y}_{\text{f}}=%
{\textstyle\sum\nolimits_{k}}
p_{k}\mathbf{Y}_{\text{f}k}$. In EQ, we keep $\mathbf{y}_{0\text{f}}%
\subset\mathbf{y}_{0}=(T_{0},P_{0},\cdots,\mathbf{A}_{0}=0)$ fixed in
generalized ensembles considered by GGR for which they derive the EQ
generalized Boltzmann distributions. As said above, NEQ ensembles require
$\mathbf{y}_{\text{f}}\subset\mathbf{y}=(T,P,\cdots,\mathbf{A})$ fixed, which
equals $\mathbf{y}_{0}$ only in EQ.

We maximize $\mathcal{S}(\mathbf{Z},t)$ to obtain $S(\mathbf{Z})=\mathcal{S}%
(\mathbf{Z})$\emph{ }under the constraint of fixed $\mathbf{Y}_{\text{f}}$ by
using the Lagrange multiplier $\boldsymbol{\lambda}_{\text{S}}%
\boldsymbol{\doteq}\partial\mathcal{S}/\partial\mathbf{Y}_{\text{f}}$
\cite[see Sec. 6.2]{Guj-entropy}; the $p_{k}$'s in IEQS are
\begin{equation}
p_{k}=\exp[-\lambda-\boldsymbol{\lambda}_{\text{S}}\cdot\mathbf{Y}_{\text{f}%
k}],\label{MicroProb}%
\end{equation}
where $\lambda$ ensures normalization; there is no explicit time dependence.
We find that $\mathcal{S=}\lambda+\boldsymbol{\lambda}_{\text{S}}%
\cdot\mathbf{Y}_{\text{f}}$ and $d\mathcal{S=}\boldsymbol{\lambda}_{\text{S}%
}\cdot d\mathbf{Y}_{\text{f}}$. Equating $d\mathcal{S}$ with $dS$ in Eq.
(\ref{Gibbs-S}), we identify $\boldsymbol{\lambda}_{\text{S}}$ with
$\boldsymbol{\lambda}$. For EQ ensembles, we obtain the result given by GGR,
which justifies the reason for this Comment.

We should mention that the maximization is not necessary for the derivation of
$p_{k}$. We can instead use the fact that being a state function
$S(\mathbf{Z})=\mathcal{S}(\mathbf{Z})$ has a unique value in the IEQ state.
Note that GGR also take $S_{\text{eq}}$\ to be a state function
\cite{Roitberg}. As $S$ is extensive and is the average of $(-\ln p_{k})$,
which is a linear operation, $\ln p_{k}$ must be a linear function of all
extensive quantities in $\mathfrak{m}_{k}$ so that we can express it as $\ln
p_{k}=\alpha+\boldsymbol{\beta}\cdot\mathbf{Y}_{\text{f}k}$ with constants
$\alpha$ and $\boldsymbol{\beta}$ so that $S=-\alpha-\boldsymbol{\beta}%
\cdot\mathbf{Y}_{\text{f}}$. Comparing $dS$ with Eq. (\ref{Gibbs-S}) gives us
Eq. (\ref{MicroProb}) with $\boldsymbol{\lambda}_{\text{S}}%
=\boldsymbol{\lambda}$. This is a much simpler derivation than the
mathematical one given by GGR.

We now make the following observation. The Gibbs-Shannon entropy
$\mathcal{S}\geq0$ always has a unique value with its minimum occurring at
$\mathcal{S}=0$. On the other hand, thermodynamic entropy\ $S$ is defined up
to a constant so its minimum is not unique. One must use the third law
($S_{\text{eq}}\overset{T_{0}\rightarrow0}{\rightarrow}0$) to fix this
constant. In that case, $\mathcal{S}$ and $S_{\text{eq}}$ can be equated
(Postulate 2) as GGR do. But there are many examples of negative $S$ (such as
an ideal gas) or of EQ crystals such as ice that have nonzero residual entropy
at absolute zero \cite[p. 467]{Pauling}. In this case, $\mathcal{S}\neq
S_{\text{eq}}$. A residual entropy also occurs in NEQ systems such as glasses,
and can be handled by modifying $\mathcal{S}$\ as shown elsewhere
\cite{Guj-entropy,Guj-entropy-2018} to ensure $S(\mathbf{Z})=\mathcal{S}%
(\mathbf{Z})$ (Postulate 2).

We end the Comment by a simple NEQ example. Consider a composite isolated
system $\Sigma$ consisting of two identical\ subsystems $\Sigma_{1}$ and
$\Sigma_{2}$ of identical volumes and numbers of particles but at different
temperatures $T_{1}$ and $T_{2}$ at any time $t\leq\tau_{\text{eq}}$ before EQ
is reached at $t=\tau_{\text{eq}}$ so the subsystems have different
time-dependent energies $E_{1}$ and $E_{2}$, respectively. We assume a
diathermal wall separating $\Sigma_{1}$ and $\Sigma_{2}$. Treating each
subsystem in EQ at each $t$, we write their entropies as $S_{1\text{eq}}%
(E_{1},V/2,N/2)$ and $S_{2\text{eq}}(E_{2},V/2,N/2)$. The entropy $S$ of
$\Sigma$ is a function of $n=4$: $E_{1},E_{2},V$, and $N$. Obviously, $\Sigma$
is in an NEQ macrostate at each $t<\tau_{\text{eq}}$. From $E_{1}$ and $E_{2}%
$, we form two independent combinations $E=E_{1}+E_{2}=$constant and
$\xi=E_{1}-E_{2}$ so that we can express the entropy as $S(E,V,N,\xi)$. Here,
$\xi$ plays the role of an internal variable, which continues to
relax\ towards zero as $\Sigma$ approaches EQ. We assume $\Sigma$ to be in IEQ
at each $t$. From $1/T=\partial S/\partial E$ and $A/T=\partial S/\partial\xi
$, we find that $T=2T_{1}T_{2}/(T_{1}+T_{2})$\ and $A=(T_{2}-T_{1}%
)/(T_{1}+T_{2})$. As EQ is attained, $T\rightarrow T_{\text{eq}}$, the EQ
temperature of both subsystems and $A\rightarrow A_{0}=0$ as expected. In this
example, $V$ and $N$ form $\mathbf{Z}_{\text{f}}^{E}$, and $\mathbf{y}%
_{\text{f}}$ refers to $T$ and $A$. The microstates $\mathsf{m}_{k_{1}}$and
$\mathsf{m}_{k_{2}}$ of the two subsystems form the microstate $\mathsf{m}%
_{k}$ of $\Sigma$. Thus, $E_{k}=E_{1k_{1}}+E_{2k_{2}}$ and $\xi_{k}=E_{1k_{1}%
}-E_{2k_{2}}$ so that
\[
p_{k}(t)=\exp[-(\lambda+E_{k}+A\xi_{k})/T]
\]
is the NEQ probability of the microstate $\mathsf{m}_{k}$, which is consistent
with the form in Eq. (\ref{MicroProb}).

Discussion with GGR is gratefully acknowledged.


\begin{thebibliography}{99}                                                                                               %


\bibitem {Roitberg}X. Gao, E. Gallicchio and A.E. Roitberg, J. Chem. Phys.
\textbf{151}, 034113 (2019).

\bibitem {Landau}L.D.\ Landau, E.M. Lifshitz, \textit{Statistical Physics},
Vol. 1, Third Edition, Pergamon Press, Oxford (1986).

\bibitem {Guggenheim}E.A. Guggenheim, J. Chem. Phys. \textbf{7}, 103 (1939).

\bibitem {Note}See Eq. (7.9) in \cite{Landau}, which shows how this entropy
formulation emerges in statistical physics. It is applicable to both EQ and
NEQ macrostates as is clear from Sec. 40 (see Eq. (40.7) in particular)
dealing with NEQ ideal gas, which we discuss later in the text.

\bibitem {Guj-I}P.D. Gujrati, Phys. Rev. E \textbf{81}, 051130 (2010).

\bibitem {Guj-entropy}P.D. Gujrati, Entropy \textbf{17}, 710 (2015).

\bibitem {Guj-II}P.D. Gujrati, Phys. Rev. E \textbf{85}, 041128 (2012).

\bibitem {deGroot}S.R. de Groot and P. Mazur, \textit{Nonequilibrium
Thermodynamics}\textbf{, }First Edition, Dover, New York (1984).

\bibitem {Prigogine}D. Kondepudi and I. Prigogine, \textit{Modern
Thermodynamics}, John Wiley and Sons, West Sussex (1998).

\bibitem {Maugin}G.A. Maugin, \textit{The Thermomechanics of Nonlinear
Irreversible Behaviors:\ An Introduction}, World Scientific, Singapore (1999).

\bibitem {Pauling}L. Pauling, \textit{The Nature of the Chemical Bond},
Cornell University Press, Ithaca (1960).

\bibitem {Guj-entropy-2018}P.D. Gujrati, Entropy \textbf{20}, 149 (2018).
\end{thebibliography}
\end{document}